# Title: Inverse-Designed Stretchable Metalens with Tunable Focal Distance

Francois Callewaert, Vesselin Velev, Shizhou Jiang, Alan Varteres Sahakian, Prem Kumar, Koray Aydin

Department of Electrical Engineering and Computer Science, Northwestern University Evanston IL 60208 USA.

**Abstract:** In this paper we present an inverse-designed 3D-printed all-dielectric stretchable millimeter wave metalens with a tunable focal distance. Computational inverse-design method is used to design a flat metalens made of disconnected building polymer blocks with complex shapes, as opposed to conventional monolithic lenses. Proposed metalens provides better performance than a conventional Fresnel lens, using lesser amount of material and enabling larger focal distance tunability. The metalens is fabricated using a commercial 3D-printer and attached to a stretchable platform. Measurements and simulations show that focal distance can be tuned by a factor of 4 with a stretching factor of only 75%, a nearly diffraction-limited focal spot, and with a 70% focusing efficiency. The proposed platform can be extended for design and fabrication of multiple electromagnetic devices working from visible to microwave radiation depending on scaling of the devices.

1.  **Introduction**

Conventional refractive optical elements based on ray optics such as lenses are typically bulky devices designed at a scale much larger than the wavelength. Recently, metamaterials and metasurfaces[1-5] have allowed the design of diffraction-based flat devices to replicate the functionalities of conventional devices with sub-wavelength or few-wavelength thicknesses. However, most metasurfaces rely on high-index building blocks assembled in a semi-periodic manner over large areas[6] (100s or 1000s wavelengths), which are well adapted for visible and near-infrared wavelengths, but too large for devices designed for longer wavelengths such as millimeter waves or microwaves. On the other hand, a computational inverse-design method has been developed recently that allows the design of on-chip photonic devices[7] and metasurfaces exhibiting interesting electromagnetic properties such as metagratings[8], metalenses[9,10], or spectral splitting[11]. Recently, we reported inverse-designed millimeter-wave flat metadevices that can bend, split, or focus plane waves over a broad range of wavelengths[12]. Here, we utilize inverse electromagnetic design[13] for the demonstration of an all-dielectric, flat and stretchable, polymer-based metalens. The lens is fabricated by 3D-printing[14] and tested in the millimeter-wave regime, showing high tunability and superior performance when compared to a Fresnel lens.

2.  **Design**

Unlike traditional methods of solving Maxwell's equations[15], inverse-design methods set all input and output boundary conditions and consider the wave equation inside the design space as an optimization problem for both the field **H** (for transversal electric waves) and the material permittivity $\varepsilon$, here for $\mu = 1$:

$$\min_{\varepsilon,H} \left\| \nabla \times \frac{1}{\varepsilon} \nabla \times H - w^2 H \right\|. \tag{1}$$

Recently, inverse-design algorithms have been used for the design of multiple photonic devices such as on-chip multiplexers[16,17], optical diodes[13], photonic crystals[18], and cloaking[19,20]. The distinctions between the design of these very different devices are the boundary conditions for both the fields and structure, initial conditions for the structure, and limits for the permittivity inside the design space, as illustrated schematically in Figure 1a. For the inverse-design of a lens, we set a constant input boundary condition and an output condition with hyperbolic phase profile[21] such that:

$$\phi(y) = \frac{2\pi}{\lambda} f \left[ \sqrt{1 + \left(\frac{y}{f}\right)^2} - 1 \right] (\text{modulo } 2\pi). \tag{2}$$

Here, the flat lens is optimized to operate for a TE polarized incoming plane wave. The focal length is chosen to be $7.5\lambda$ at the optimized wavelength without stretching. Once the boundary conditions are set, the degrees of freedom of the inverse-design algorithm are the range of permittivities allowed and the size and initial conditions of the design space. The range of relative permittivities is defined by the materials used for the fabrication, here air, $\varepsilon = 1$ and high impact polystyrene (HIPS), $\varepsilon = 2.3$.

In order to generate the hyperbolic output phase profile in Eq. (2), the design space must be large enough to allow phase change variations of up to $2\pi$ between a part full of air and a part full of polymer. Heuristically, this condition is achieved with a lens thickness as low as $1.5\lambda$, approximately, which is the thickness that we choose for the optimization. Finally, we set the initial condition of the optimization routine to a uniform permittivity level $\varepsilon_{init}$. This parameter has a considerable influence on the optimized design structure and performance[13]. Depending on its value, the final device will be either mostly made of polymer (when $\varepsilon_{init} = 2.3$) or made of sparse blocks (when $\varepsilon_{init} = 1$), as illustrated in Figure 1b, where we show three structures obtained via inverse-design optimization with initial permittivities of 2.3, 1.65, and 1.0 respectively. With seven individual blocks and a polymer content of only 30% of the design-space area, the third device is ideal to build a stretchable lens. This device can be either compressed by a factor down to $s = 0.7$, or stretched by any factor $s > 1$, as illustrated in Figure 2, where structures compressed by a factor of 0.8, 1.0, and 1.4 are shown.

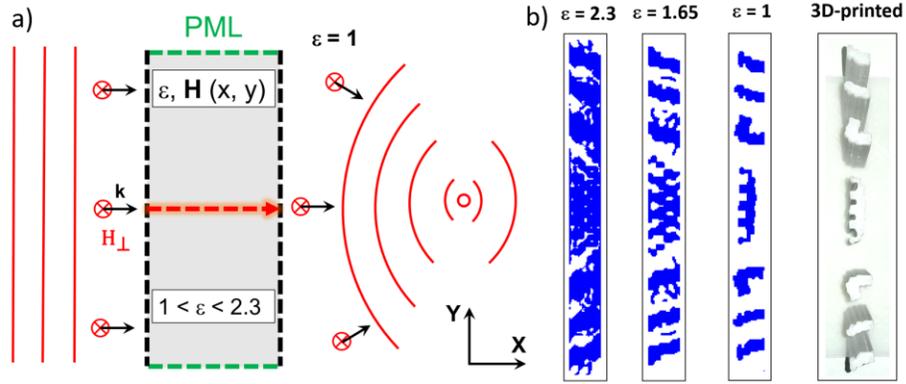

**Figure 1:** a) Schematic representation of the inverse design of a lens. A TE-polarized input plane-wave perpendicularly incident to the left of the design space undergoes a transformation in the device to become an output cylindrical wave focusing at a desired focal distance. The algorithm optimizes both the permittivity and the perpendicular magnetic field inside the design space. PML = perfectly matched layer. b) The left three images show structures optimized by the design algorithm starting from initial uniform permittivities of $\varepsilon = 2.3$, 1.65, and 1.0, respectively. The final image is a photograph of the 3D-printed HIPS lens from the $\varepsilon = 1$ simulation.

## 3. Experiment

The device was printed with a commercial 3D-printer, and a picture is shown in Figure 1b. The device's $X \times Y$ dimensions are $1.5 \times 15\lambda$, which corresponds to $1.25 \times 12.5$ cm for an operating wavelength of 8.3mm (36GHz). For such a 2D lens, the height is simulated as infinite, and in practice the device is 10cm-thick (12$\lambda$), which takes around 10 hours to print. After printing, we connect all seven blocks with each other at the top and bottom with two rubber bands, which act like a stretchable platform. When the rubber bands are quiescent, the device has a stretch factor of $s = 0.8$, for a length of 10cm. The device can then be stretched by a factor up to $s = 1.5$. We test the device's response to normally-incident electromagnetic radiation by using a radiofrequency source to generate a millimeter-wave beam through a high-gain horn antenna which directs the radiation perpendicularly towards the device. The device is placed 1m away from the antenna so that the beam is approximately a plane wave, and the output power is mapped along a plane on the other side of the device using a WR-28 waveguide (3.5×7mm) attached to a X-Y stage. The measurement starts around 1cm to the right of the device due to technical limitations of the setup. We measured the output power for stretching factors of $s = 0.8, 1.0, 1.2,$ and $1.4$, which corresponds to lengths of 10cm, 12.5cm, 15cm, and 17.5cm, respectively. We also simulated the electromagnetic behavior of the device with full-field FDTD simulations. In Figure 2, we show the map of the power profile along the axial plane at 36GHz from simulations (left) and experiments (right) for stretching factors of 0.8 (top), 1.0 (middle), and 1.4 (bottom). As can be seen, there is a remarkable agreement between the simulations and the experiment, showing that the device behaves as a tunable lens as expected.

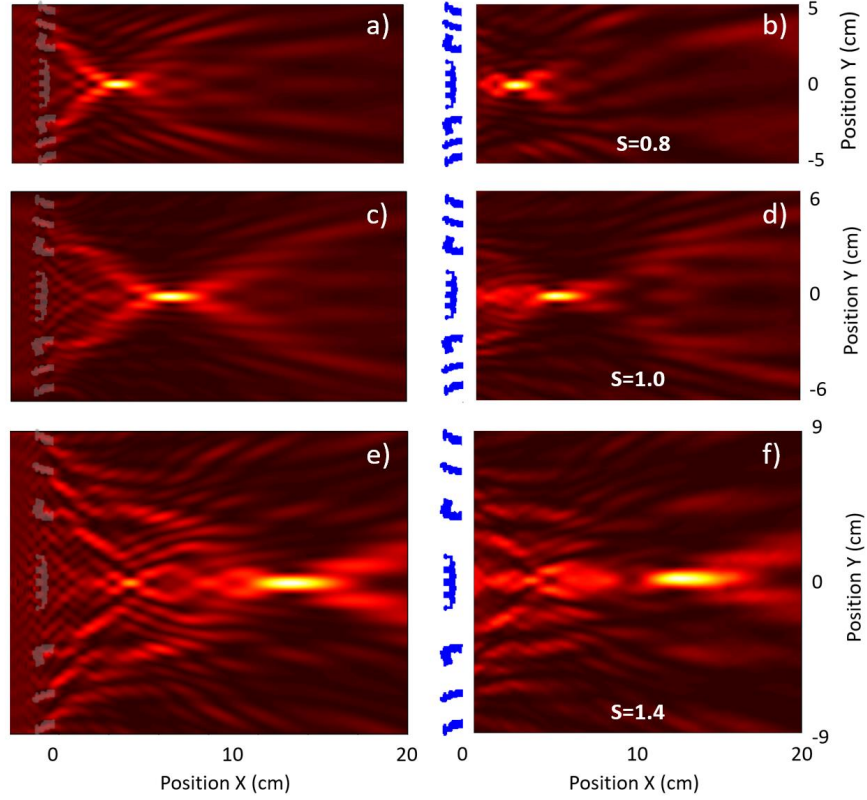

**Figure 2:** Simulated (left) and experimental (right) optical power profiles along the axial plane of the device for stretching factors of $s = 0.8$ (a and b), 1.0 (c and d), and 1.4 (e and f).

We report the focal distance in Figure 3a for all stretching factors and for frequencies from 30GHz ($\lambda$ = 10mm) to 40GHz ($\lambda$ = 7.5mm). The focal distance can be fitted with the following heuristic law:

$$f \approx \left(\frac{\lambda_0}{\lambda}\right)^{\frac{3}{2}} \left(s^2 f_0 + s(s-1)\frac{L}{6}\right), \qquad (3)$$

where $f$ is the focal distance, $f_0$ is the focal distance at optimal wavelength $\lambda_0$ and at stretching factor $s = 1$, and $L$ is the lens length when $s = 1$. The first term in the parenthesis comes from the paraxial approximation, which is represented by the dashed line in Figure 3a, and the second term is a correction due to the high numerical aperture of the lens, which has magnitude comparable to the focal distance. Experimentally, we observe that the focal distance is tuned from 4.5cm to 17cm for a stretching factor from 0.8 to 1.4, which is tuning of a factor of 3.8 for a relative stretching factor of 1.75.

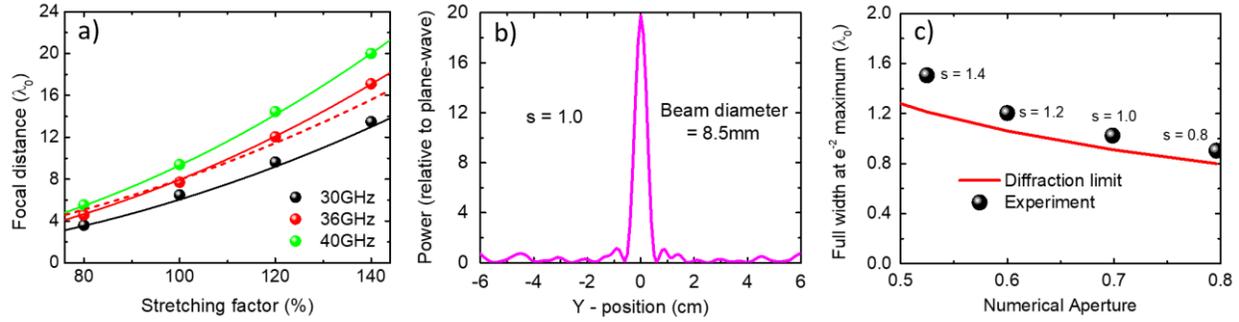

**Figure 3:** a) Experimental (spheres) and theoretical (solid lines) focal distance as a function of the stretching factor and the frequency, expressed as a factor of the optimal wavelength $\lambda_0$. The dashed line represents the theoretical distance in the paraxial approximation. b) Experimental power profile in the focal plane of the device for a stretching factor $s = 1.0$ and a frequency of 36GHz ($\lambda_0$ = 8.3mm). The power is normalized to the power of the plane-wave reaching the device. c) Measured and diffraction-limited beam spot size in the focal plane as a function of the numerical aperture of the lens, which increases with lower stretching factors.

In Figure 3b, the measured power profile is plotted in the focal plane of the device for $s = 1.0$ and at 36GHz, normalized by the power of the incident plane wave. The power at the focal point is 17 times higher than the power in any other point of the focal plane. The beam diameter, defined as the full width at $e^{-2}$ ($\approx 0.135$) maximum is 8.5mm, which is very close to the diffraction-limited value of $D = 7.6$mm. The beam diameter is reported in Figure 3c as a function of the numerical aperture for all four experimental stretching factors and compared to the diffraction limit, given by:

$$D = \frac{2\lambda}{\pi\,\mathrm{NA}}, \tag{4}$$

where NA is the numerical aperture, equal to 0.7 for a stretching factor $s = 1.0$, and which follows:

$$\text{NA} = \left[1 + \left(\frac{f(s)}{sR}\right)^2\right]^{-\frac{1}{2}}. \qquad (5)$$

The lowest NA corresponds to the highest stretching factor ($s = 1.4$) and the highest NA to $s = 0.8$. As can be seen, the stretchable lens is very close to being diffraction limited for $0.8 < s < 1.2$, and the experimental beam diameter for $s = 1.4$ is still only 24% higher than the diffraction-limited value.

## 4. Comparison with Fresnel lens

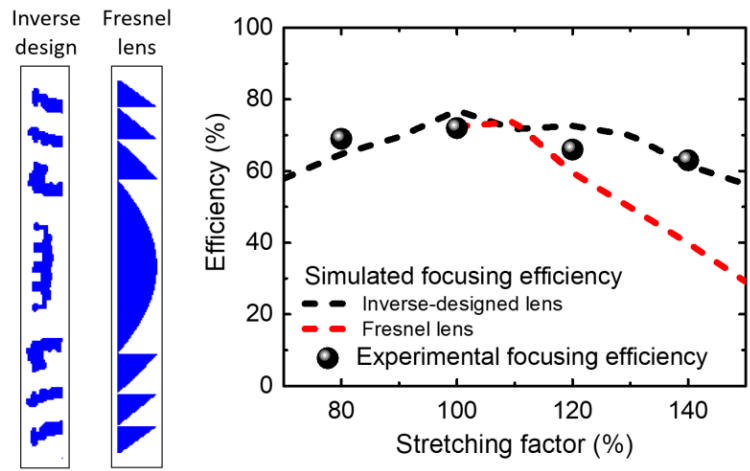

**Figure 4:** (Left) Comparison between a Fresnel lens and the inverse-designed device, showing that the latter uses only 44% as much material. (Right) Experimental (spheres) and simulated (dashed lines) focusing efficiency of the inverse-designed device and the Fresnel lens as a function of the stretching factor. Focusing efficiency is defined as the ratio of the power going through the center peak divided by the power transmitted through the focal plane.

A Fresnel lens with the same refractive index, matching size, focal distance, and operating wavelength is shown in Figure 4 and compared to the inverse-designed device. Both devices rely on diffraction and can be viewed as first-order gratings with periodicity that decreases farther from the center. However, the inverse-designed device has the advantage of using only 44% as much material as the Fresnel lens, and it can be uniformly compressed by a factor as low as $s = 0.7$ for better tunability, which is not possible with the Fresnel lens. We simulated the focusing efficiencies of both devices at the optimal wavelength and for stretching factors from 0.7 to 1.5 (1.0 to 1.5 for the Fresnel lens) and plotted them in Figure 4 (dashed lines) to compare their performance. Focusing efficiency is defined here as the ratio of the power that passes through an aperture in the focal plane with a size of the beam diameter as defined earlier, over the total power going through the focal plane. As can be seen, both devices have similar performance $\approx 75\%$ at low stretching factors, but the inverse-designed device has better efficiency for larger stretching factors, with efficiency $> 60\%$ for $0.7 < s < 1.4$, compared to only $1.0 < s < 1.2$ for the Fresnel lens. This is explained by noting that the focal point corresponds to the first-order diffraction mode of the lens-grating, and power tends to be scattered into higher-order modes when the Fresnel lens is stretched. Experimental focusing efficiency of the inverse-designed device matches closely with the simulated efficiency and remains within 63% and 72% over the entire range of stretching factors.

## 5.     Conclusion

In conclusion, we have designed and experimentally demonstrated a low-index flat and stretchable lens in the millimeter-wave regime. By carefully selecting the initial condition, the inverse-design algorithm converged towards a structure made of seven distinct building blocks. The device can then be stretched or compressed to adjust the focal distance from $4.7\lambda$ to $17\lambda$ with a relative stretching of 75% only, while keeping focusing close to diffraction-limited and with high focusing efficiency around 70%. Experimental results agree remarkably well with the simulations for the entire range of stretching considered, which shows the effectiveness of the proposed inverse-design and 3D-printing platform for the design and fabrication of efficient electromagnetic devices. The method could be used for the design and fabrication of devices with various properties such as lensing, polarization sensing, or holograms. Additionally, thanks to the very stable dielectric properties of polymers[22,23] and using the wide range of 3D-printing methods available[14,24], any design can be scaled to operate from the microwaves to the visible.

**Acknowledgements**

K.A. acknowledges support from the Office of Naval Research Young Investigator Program (ONR-YIP) Award.